\documentclass[]{spie}  
\usepackage[]{graphicx}

\title{The opto-mechanical alignment procedure of the VLT Survey Telescope}

\author{Carmelo Arcidiacono\supit{a,b} Roberto Ragazzoni\supit{b} Gabriele Umbriaco\supit{c} Jacopo Farinato\supit{b} and Demetrio Magrin\supit{b}
\skiplinehalf
\supit{a}INAF-Osservatorio Astrofisico di Arcetri, Largo Enrico Fermi, 5, Firenze, Italy; \\
\supit{b}INAF-Osservatorio Astronomico di Padova, Vicolo dell'Osservatorio, 5, Padova, Italy; \\
\supit{c}Universit\`a degli studi di Padova, Vicolo dell'Osservatorio, 2, Padova, Italy; \\
}

\authorinfo{Further author information: (Send correspondence to Carmelo Arcidiacono)\\C.A: E-mail: carmelo@arcetri.astro.it, Telephone: +39 055 2752 293}

\begin{document}
  \maketitle

\begin{abstract}
The VLT Survey Telescope is a f/5.5 modified Ritchey-Chretien imaging telescope, which is being installed at the ESO-Paranal
Observatory. It will provide a one square degree corrected field of view to perform survey-projects in the wavelength range
from UV to I band. In this paper we describe the opto-mechanical alignment procedure of the 2.61m primary mirror, the
secondary and correctors lenses onto the mechanical structure of the telescope. The alignment procedure does not rely on the
mechanical precision of the mirrors. It will be achieved using ad-hoc alignment tools, described in the paper, which allows the
spatial determination of optical axes (and focuses where necessary) of the optical components with respect to the axis defined
by the rotation of a laser beam mounted on the instrument bearing.
\end{abstract}


\keywords{Telescope, optical alignment, VST}

\section{INTRODUCTION}
\label{sec:intro}  
The VLT Survey Telescope (VST) project is a cooperation between
the European Southern Observatory
(ESO) and the Italian National
Institute of Astrophysics (INAF) for the
design and realization of a wide field Alt-Az
telescope of 2.61-m aperture, specialized
for high quality astronomical imaging,
to be installed and operated at the ESO
Paranal Observatory in Chile close to the
VLT units.

In this paper we briefly describe the optical\cite{2005SPIE.5962..608M} and mechanical\cite{2003SPIE.4837..379M} design of the VST telescope in order to have all the elements to finally describe the alignment procedure
we planned to follow for the mechanical and optical components of the telescope.

\section{VST Design}
The VST is a 2.61 m Alt-az f/5.5 modified Ritchey-Chretien telescope with a corrected Cassegrain Field of View (FoV) diameter of 1.47degrees,
coupled with the OmegaCAM\cite{2002Msngr.110...15K} instrument, a 16k$\times$16k CCD mosaic
camera of 15$\mu$m pixel size and 0.21arcsec/pixel scale.
 In order to cover the wide field of view with the required high image quality, the telescope is not a pure Ritchey-Chretien, but the design includes two different camera correctors respectively with and without dispersing
elements to compensate atmospheric dispersion Also the OmegaCAM dewar window is a low power lens
itself. 
One corrector camera is made of two lenses (L1 and L2 hereafter) and operates from U to I bands. The
other one is made with an atmospheric dispersion corrector (ADC) and one different lens (L3) to observe from B to I bands. The two correctors can be exchanged by mean of a large translator stage switching the two systems. The telescope is provided with active optics both for secondary and primary mirrors in order to compensate for telescope
structure deformations.

The telescope opto-mechanical design includes an instrument interface flange, which divides the
telescope in two blocks along the optical axis. The first block includes mirrors and primary mirror cell with correctors and a pick-up
mirror for the active optics (Shack-Hartmann) sensor and the second block includes filters, dewar window lens and the OmegaCAM instrument camera.
The secondary mirror is mounted on a dedicated high resolution positioning system. This system is
a double stage hexapod device which guarantees micro metric and sub-micro metric
displacements and high resolution tilts of the secondary mirror (M2) with respect to the primary (M1).

The final goal of the alignment procedure is to place the M1, M2 and the two correctors aligned along the optical axis and at the proper distances within the range of adjustment offered by the active optics system.



\section{Telescope Alignment}
\label{sec:align}
The alignment procedure was inspired by the one used to align the mirrors of the Large Binocular Telescope (LBT) and the LBT prime focus Camera (LBC)\cite{2004SPIE.5492..513D,2008SPIE.7014E.159G}.
We will use an ad-hoc alignment tool, built by Tomelleri S.r.l., to align the optical components of the VST telescope. This is a pure optical procedure, which does not trust in the precision of any mechanical device, such as in other possible alignment procedure followed for the Telescopio Nazionale Galileo (TNG) and the Very Large Telescope (VLT).
We designed an ad-hoc device (an ``alignment tool'') for this task which will be used for the M1 and M2 optical alignment.
\subsection{Mechanical Alignment}
\begin{figure}
   \begin{center}
   \begin{tabular}{c}
   \includegraphics[width=8cm]{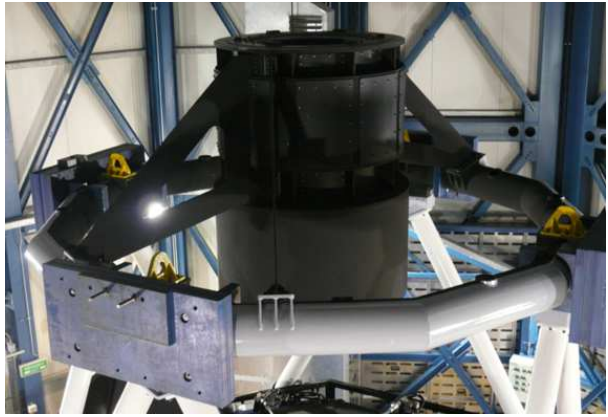}
   \end{tabular}
   \end{center}
   \caption[fig1]
   { \label{fig:fig1}
The figure shows the M2 cell and the spider.}
   \end{figure}

A pre-requisite for the optical alignment is the mechanical alignment of the altitude and azimuthal axes with respect to the OmegaCAM instrument bearing that we will use as reference for the optical procedure.
In order to have a good starting position that allows to have most of the dynamic range available on the M2 hexapod we plan to shim the spider of M2 to center it with respect to the mechanical axis of the telescopes. The spider is mounted using flanges mounted at $\approx$45degrees (Figure~\ref{fig:fig1}), considering that this shimming will introduce extra Z axis shift of the same order of the decentering introduced.
Similar shimming will be applied to the primary mirror cell, but only for tilt, being installed for the centering of the M1 mirror three translation pads mounted on screws.

The reference for the alignment of the M1 cell and secondary mirror spider will be a sighting telescope mounted below the azimuth axis on the basement of the VST dome (the interface is already installed).
In this way we can also define the encoder Zenith value for the elevation movement of the telescope, which places on the same plane the two axes defining the Azimuth and the OmegaCAM rotation bearing. Such as angle will be taken into account in the pointing model of the telescope.
First we will adjust the M2 mounting (using the target mounted on the M2 cell) shimming eventually the spider, then we place the autocollimator target on the instrument bearing and we will adjust the cell tilt mechanically. Such as an operation will be repeated at the very end of the alignment procedure.
The alignment of the bearing with respect to the azimuth axis will prevent field rotation issues because if bearing axis is different with respect to the telescope pointing movements  it will have a different centre of rotation. Please notice that a tilt of the bearing with respect to Azimuth generates a misalignment for the field correction derotation, while a decenter will generated only a different speed for derotation visible in long exposures images.
\subsubsection{Operations}
\begin{figure}
   \begin{center}
   \begin{tabular}{c}
   \includegraphics[width=8cm]{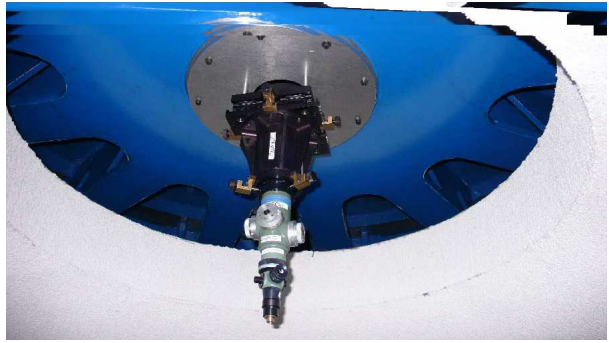}
   \end{tabular}
   \end{center}
   \caption[fig2]
   { \label{fig:fig2}
The figure shows the sighting telescope ``STaz'' mounted below the VST azimuth axis with a special flange.}
   \end{figure}

The sighting telescope will be mounted on the fixed platform below the VST telescope (see Figure~\ref{fig:fig2}) and the target mounted on the azimuth flange. The sighting telescope will be aligned with respect to the azimuth axis by rotating it.
A second target mounted on the OmegaCAM bearing will be adjusted in centering by rotating the bearing itself with respect to the axis defined the bearing itself.
The telescope altitude axis will be moved in order to reach the closest position between the axis of the OmegaCAM bearing and the axis of the telescope azimuth. We will take note of the encoder value of the altitude motion relative to this position because this will define the Zenith position.
A third target will be mounted on the nominal center of the M2 hexapod, with the hexapod at the mid of the available range of motion, in order to measure the decenter of the M2 itself.
This configuration will allow to compute the tilt of the OmegaCAM bearing with respect to the Azimuth motion and to design the shims needed to compensate it. The shim will be inserted below the four flexion bars. This operation implies the demounting and remounting of the cell (without the OmegaCAM dummy mounted). The manufacturing of the shims will be done in the mechanical workshop of the Paranal Observatory.
The measurement relative to M2 target will provide the decentering of the M2 spider, that will be corrected by shimming the spider and/or the hexapod mounting.
After the remounting of the M1 cell and M2 hexapod and spider, the two targets will be realigned, the elevation encoder value measured again, and the decentering and tilt of the cell and the decentering of the M2 measured verified.
The M2 reference center position will be the one taking into account the flexure expected at a mid elevation altitude in order to save range to compensate the decentering of the M2 due to off the Zenith telescope elevations.

After this phase we will check the ``alignment tool'' mounting on board of the telescope: we will test the procedure to mount this tool and the one to dismount it. We strongly recommend to verify this procedure after the mechanical alignment being the last opportunity to have the M1 dummy in the cell. We will check also the wood protection of the M1 mirror to be mounted below it. This device will be attached the alignment tool to the instrument bearing interface.
\subsection{Alignment Procedure}
\begin{figure}
   \begin{center}
   \begin{tabular}{c}
   \includegraphics[width=8cm]{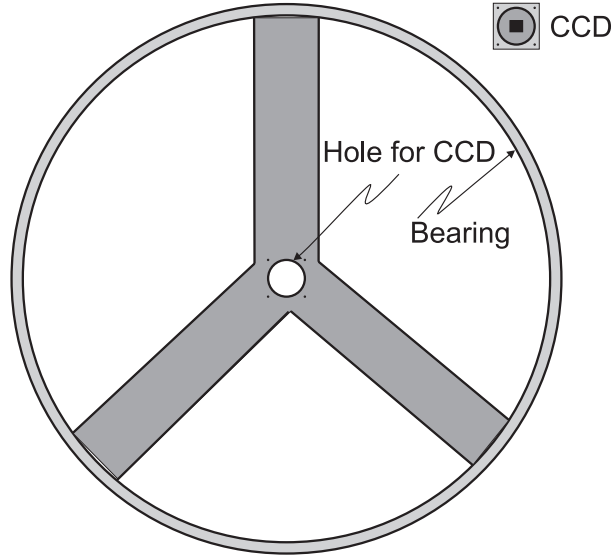}
   \end{tabular}
   \end{center}
   \caption[fig3]
   { \label{fig:fig3}
The figure shows the flange designed for the T-Tower. It is mounted on the instrument bearing. The hole let the light passing through, either for the adjustable laser beam or for the CCD looking to the light fired by the laser mounted on the horizontal flange above the M1.}
   \end{figure}

\begin{figure}
   \begin{center}
   \begin{tabular}{c}
   \includegraphics[width=12cm]{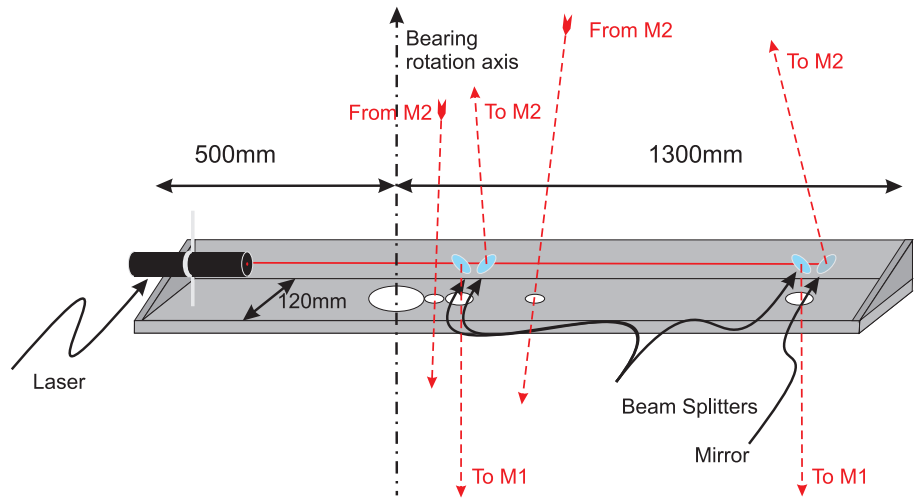}
   \end{tabular}
   \end{center}
   \caption[fig4]
   { \label{fig:fig4}
   The horizontal arm of the T-Tower, a card-board mechanism will switch between the configurations with beams going to M1 or going to M2. This arm will be mounted on a truss made in way to let the beam coming from the M2.}
   \end{figure}

As before, the reference for the alignment will be the 2$^{\rm nd}$ bearing rotation axis (the instrument one), which precision will set the limit of the final alignment quality.
The ad-hoc mechanical device is mounted on the bearing axis using a flange (see Figure~\ref{fig:fig3}) and, a mechanical vertical extension (see Figure~\ref{fig:fig6}) supports an horizontal rail (see Figure~\ref{fig:fig4}) to the area immediately above the primary mirror. This rail on a side (the short one) supports the laser and on the other, long about as the radius of the M1, supports a 2~inches folding mirror and three 50/50 beam splitters. On this rail the laser is able to fire (remotely commanded) a beam to the secondary or to the primary mirror using these folding mirrors (beam-splitters): the selection of the beam up- or down- warding can be accomplished by a moving cardboard (actually steel made) with selection holes.
Such as device will be mounted on a flange centered with respect to the optical axis, passing through the ADC corrector barrel. Of course the ADC correctors lens and prisms have to be removed from this barrel and they will be mounted only at the end of this procedure.
The lower baffling above the M1 has to be removed. This is doable without the removal of the cover (petals) of the mirror or other elements in the Serrurier truss area. However, in order to place the alignment device, the petals covering the M1 will be removed.
The alignment operation will start with the two lenses of the corrector mounted in the respective barrel, while the ADC prisms and the lens are not mounted in the ADC barrel.
During the alignment the stage mounting the corrector barrels is in the so called ``ADC'' position, in this way there is the physical room to have the ``alignment device", ``T-Tower" hereafter, going through the hole of the M1 mirror. The corrector lenses L1 and L2 can be placed in position without dismounting nothing of the telescope (for example the bearing). The placing in position of the two prisms and of L3 lens is not possible without dismounting the corresponding barrel: therefore the so called ``ADC'' stage will be dismounted from the cell and put back in position using mechanical pins as precision references.
We provide the T-Tower with an insertable detector (``CCD'' in the following) mounted on the focal plane area of the telescope. Such as device will be unavoidably subjected to flexures due to gravitational forces. However, as the telescope is pointing to the local Zenith and the only movement is around the gravity axis (rotation of the bearing) such as flexures will be invariant for the whole alignment operations.

The alignment device is mounted directly to the bearing with a ``Y" flange (see Figure~\ref{fig:fig3}) and with a vertical truss (see Figure~\ref{fig:fig6}) passing through the ADC barrel in order to match the bearing axis. The horizontal arm of the T-Tower will stay $\approx$700mm far from the M1 dish.

On top of this, about 700mm far from the M1, an horizontal flange hosts a laser: this fires a beam to the secondary or to the primary mirror. The selection of the up- or down- warding direction is done by using beam splitters and a small folding mirror. The vertical mechanical extension is a truss divided in two parts in order to let beams coming from M2 to reach the insertable CCD at the level of the focal plane.
In this way it is possible to realize both chief and marginal rays. By rotating the bearing with the laser turned-on is possible to illuminate different portions (rings) of the mirrors (M1 and M2) and adjust iteratively centering and tilt of both M1 and M2. The selection of the chief vs. marginal rays can be done through a cardboard shutter possibly remotely controlled.
   \begin{figure}
   \begin{center}
   \begin{tabular}{c}
   \includegraphics[width=8cm]{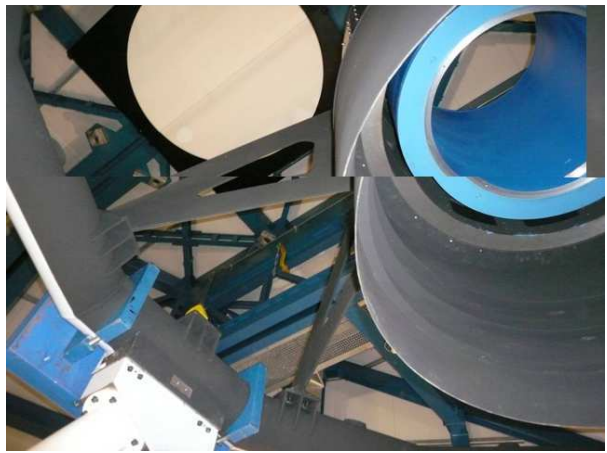}
   \end{tabular}
   \end{center}
   \caption[fig5]
   { \label{fig:fig5}
   The picture shows lower baffle that must be removed  to place in position the T-Tower. It can be removed passing through the Serrurier truss.}
   \end{figure}

   \begin{figure}
   \begin{center}
   \begin{tabular}{c}
   \includegraphics[width=8cm]{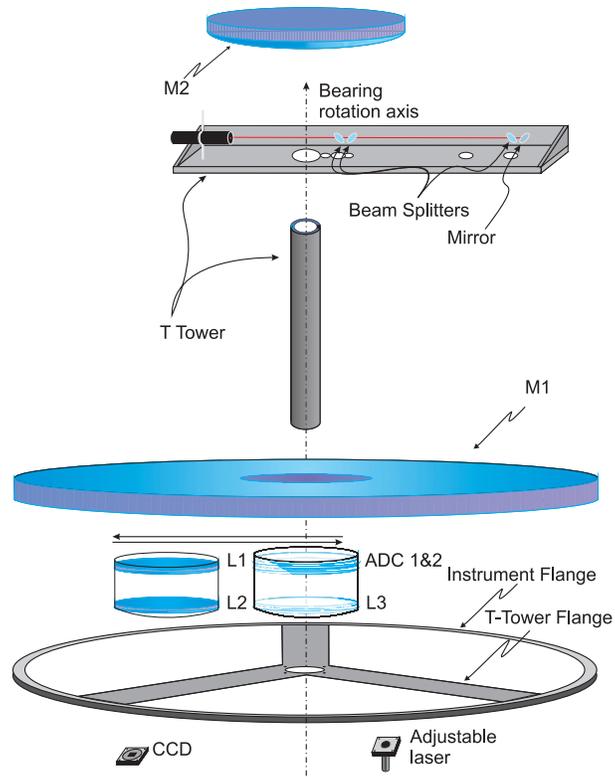}
   \end{tabular}
   \end{center}
   \caption[fig6]
   { \label{fig:fig6}
   Sketch of the components needed for the ``LBT" like optical alignment. The T-tower vertical support will pass through the empty barrel of the ADC corrector. The two ADC prisms and L3 lens will be inserted in a later stage, after the alignment of M2, M1 and Corrector (L1+L2).}
   \end{figure}

   \begin{figure}
   \begin{center}
   \begin{tabular}{c}
   \includegraphics[width=8cm]{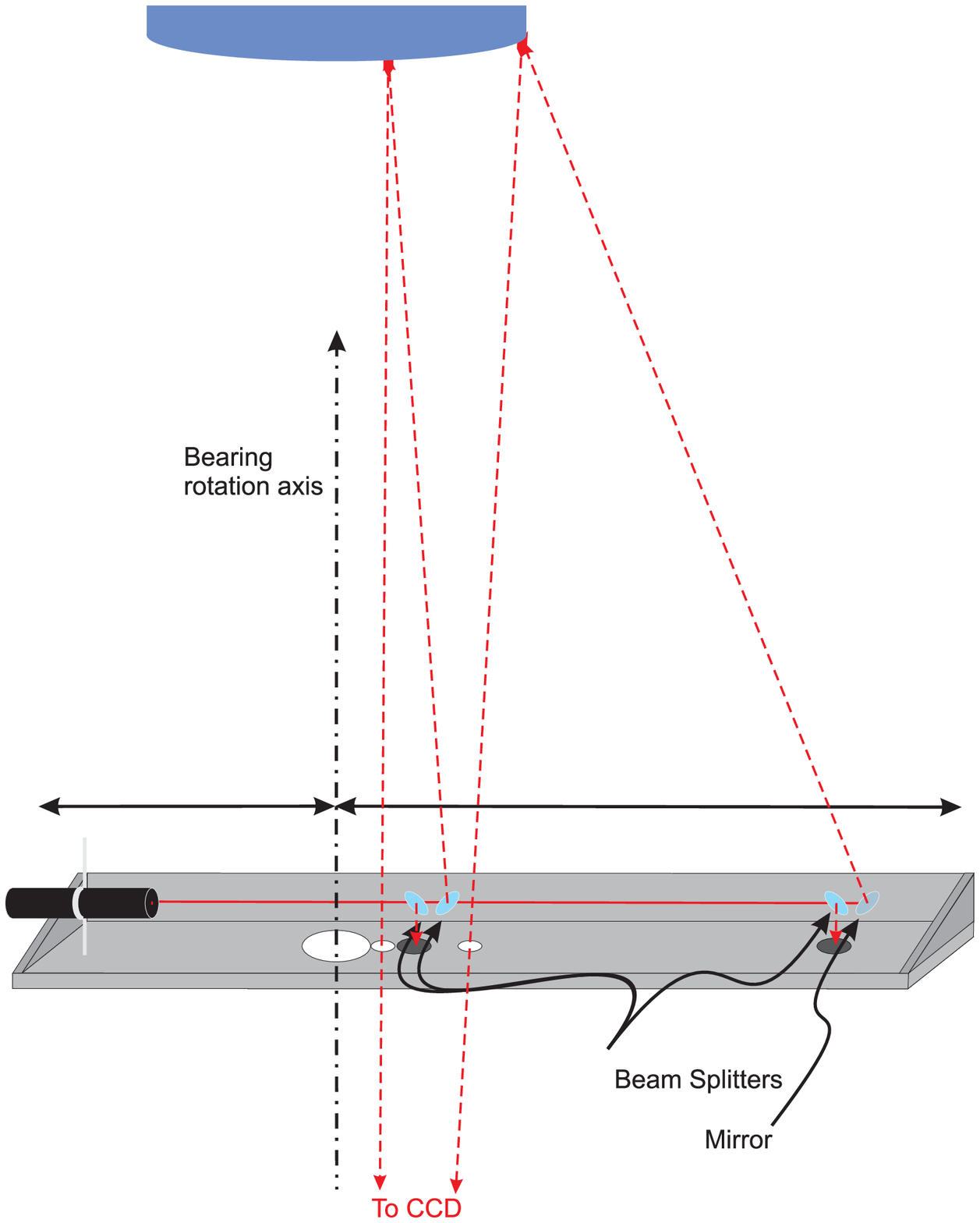}
   \end{tabular}
   \end{center}
   \caption[fig7]
   { \label{fig:fig7}
   The figure shows a sketch of the alignment of the M2 mirror.}
   \end{figure}

\subsection{M2 and M1 alignments}
The procedure will start with the M2 alignment, only after this phase M1 will be placed in the correct position. The alignment scheme of M1 and M2 is the following:
\begin{enumerate}
\item the CCD mounted at the bottom of the T-Tower is illuminated with any kind of source fixed with respect to the fixed part of the telescope. By rotation of the device the illumination pattern does apparently rotate onto the CCD and one can individuate the position on the detector that is intercepted by the bearing axis. In the following such a position is called ``center of the CCD".
We will rotate the CCD only using the bearing and illuminating it with a fixed led light: in this way we see a pattern which rotates and we can identify the rotation pivot pixel.
\item 	Using the configuration in Figure~\ref{fig:fig7}, the ``chief" ray is materialized by the laser and illuminates the secondary mirror. It is adjusted in order to match approximately the CCD center. Now the bearing is rotated and the secondary mirror is adjusted with only tilt in order not to move during the rotation;

\item	The ``marginal" ray is materialized and the bearing is rotated and the secondary is adjusted in order to have no apparent movement of the reflected spot with only rotation around the center of curvature of the osculating sphere on the M2;
\item	Steps 2 and 3 are iterated till to reach the no movements of the spot condition on the CCD during the bearing rotation;
\item	NOW the M2 is aligned with the bearing.

   \begin{figure}
   \begin{center}
   \begin{tabular}{c}
   \includegraphics[width=8cm]{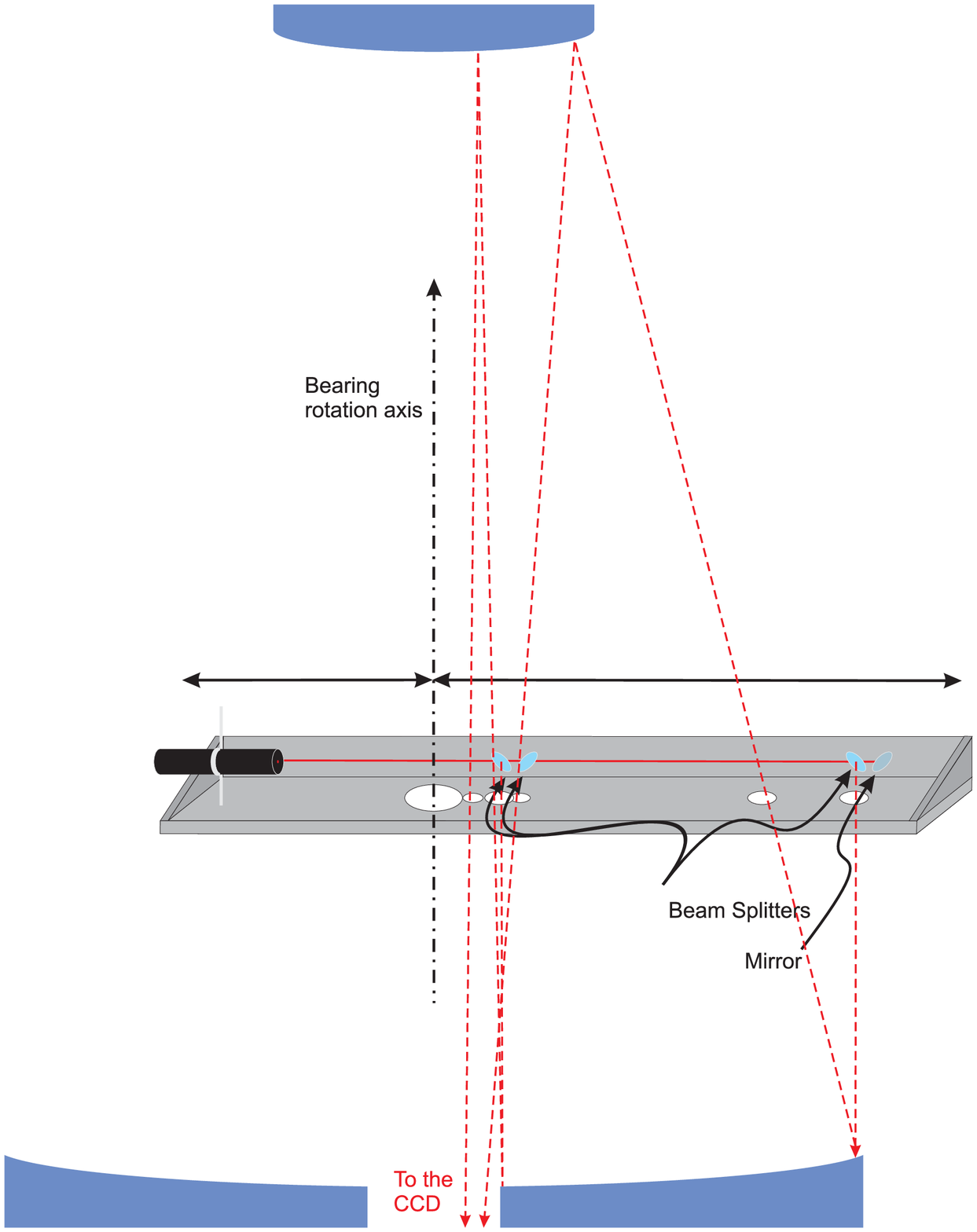}
   \end{tabular}
   \end{center}
   \caption[fig8]
   { \label{fig:fig8}
   The figure shows a sketch of the alignment of the M1 mirror: the laser beam is reflected by the M1 and M2 before reaching the CCD mounted on the Y shaped flange at the level of the expected focal plane.}
   \end{figure}

   \begin{figure}
   \begin{center}
   \begin{tabular}{c}
   \includegraphics[width=8cm]{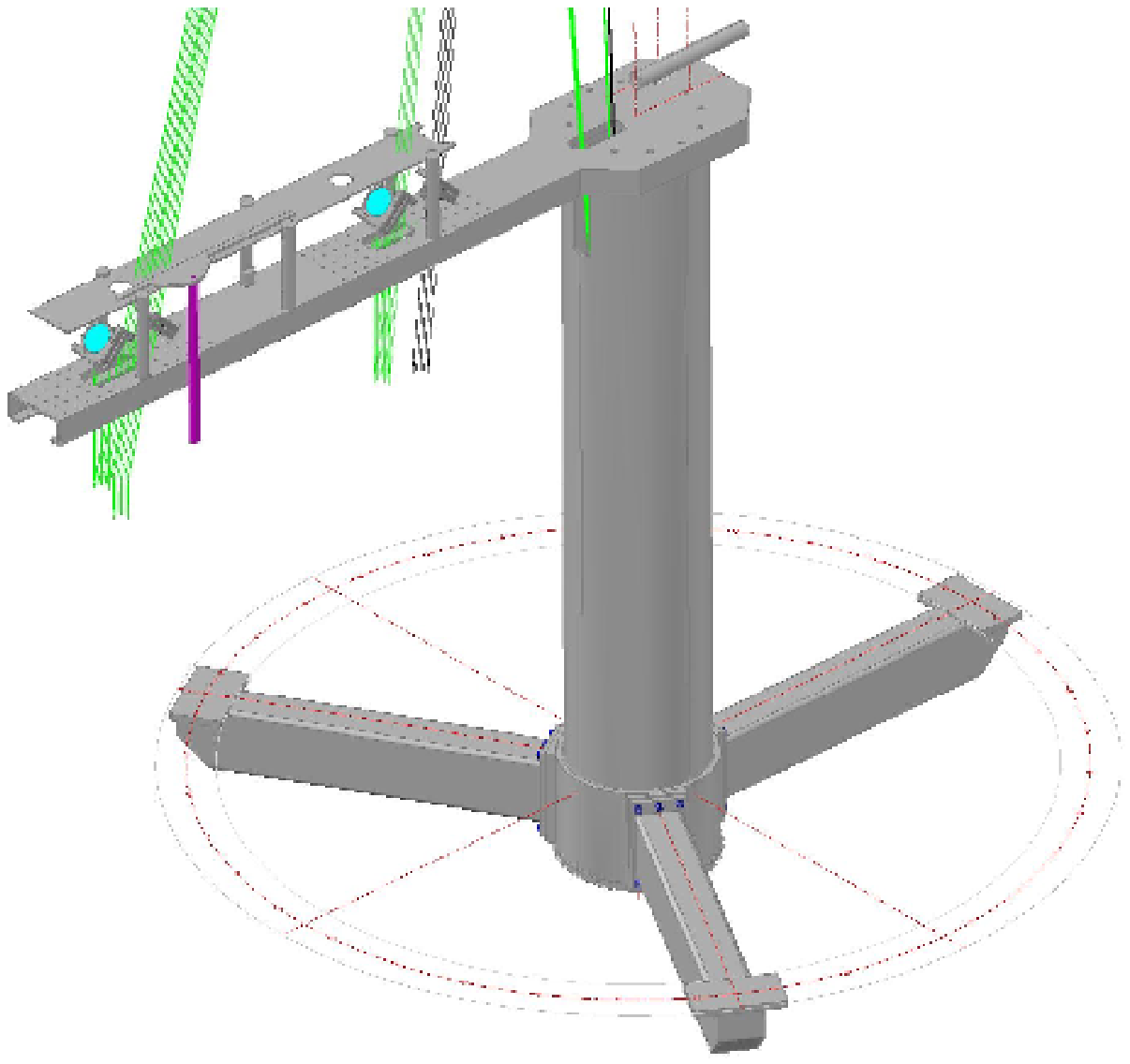}
   \end{tabular}
   \end{center}
   \caption[fig9]
   { \label{fig:fig9}
   The figure shows a CAD view of the alignment tool system T-Tower. Super-imposed to the image are the optical rays as they actually will fly from the laser reflected towards the M1, back reflected, passing through the holes made on the T-arm and directed to the M2, back reflected again and imaged on the CCD mounted on the bottom of the truss.}
   \end{figure}

\item	The laser beam is now fired toward the primary mirror, reflected by the secondary and finally reaches the CCD mounted on the T-Tower Y-flange. Procedures similar to the steps 2 and 3 are now iterated using the primary mirror adjustments. Using the chief ray we set tilt, and using the marginal one we set decentering. At the end of this level the primary mirror is also co-aligned;
\item After the alignment of M1 we plan to look to a bright star on the sky in order to refine focus and mirror position looking to the pupil image. In this case we will remove the T-tower truss, but not the CCD that will be used both as reference and as detector;
\item Still looking to a star it is possible to make out of focus images and correct coma through M2 rotation around its center of curvature thanks to a trials and errors procedure.
\end{enumerate}
\subsection{Corrector(s) alignment}
\begin{figure}
   \begin{center}
   \begin{tabular}{c}
   \includegraphics[width=8cm]{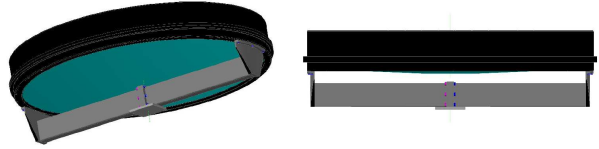}
   \end{tabular}
   \end{center}
   \caption[fig10]
   { \label{fig:fig10}
   The figure shows a CAD view of the interface to be mounted in front of M2 to place the CCD used for the definition of the optical axis.}
   \end{figure}

The insertion of the corrector lenses, using the ADC stage mechanism, is possible while the laser is in position on the alignment tool. The insertion of the L3 lens and of the two prisms inside the ADC barrel will be done in only on a later stage. We have provision for pins of the corrector mounting mechanics with respect to the M1 cell in order to remove it, insert L3 and prisms in the barrel, and place the mechanics in to its old position. The pinning was already done in Italy before the shipment of the M1 cell and correctors systems.
In principle, since the optical tolerance of the correctors are quite large (of the order of 3-5mm in centering, because of the low optical power of the lenses) there is a not negligible probability that mechanical precision is enough to have the corrector elements already aligned, therefore we planned to point on sky, and to simply place the corrector lenses barrel on axis and to have a first look at the sky in the center of the FoV: if the dimension of the imaged stars on the focal plane will be in the specification the corrector barrel can be considered aligned, otherwise we will follow the following steps.
An adjustable laser is installed on the bearing, using the ``Y" base of the T-Tower, and adjusted iteratively with the help of a close by and a far away CCDs in order to materialize the bearing (and now optical) axis.
The two CCDs for the alignment of the laser will be mounted on ad-hoc mechanical mountings:
\begin{itemize}
\item	 in front M2, mounted on a ad-hoc flange (see Figure~\ref{fig:fig10});
\item on the M1 cover;
\end{itemize}
The barrel with the corrector lenses is place in position using the ADC translation stage and the laser spot refraction and the back reflection are used to adjust the centering and tilting of the lenses. We start with the lower one (L2):
\begin{enumerate}
\item	Tilting of L1 and L2 will be done using ad-hoc 0.1mm shims on the barrel;
\item	Centering is done using the translation correctors stage and moving the whole plate, on which the stage is mounted, on the orthogonal direction (there is a provision for this movement);
\item	In the case of L1, while is not in the L2 case, light back reflected from the first of the two corrector lenses, once this is aligned, is useful, through interference to align the second one.
\item	Please notice that the back reflected beam is observable only through a holed screen and projected there, while the refracted beam can be measured through one of the CCDs used for the laser alignment.
\item	We compute the size of the shims to be inserted on L1 and L2;
\item The corrector barrel will be dismounted and the shims will be inserted on the side of L1 and L2;
\item	The corrector is mounted back;
\item   Adjustable Laser aligned as described at the step 5;
\item	L1 and L2 moved in position using the corrector stage;
\item	alignment will be verified.
\end{enumerate}
Now the two corrector lenses are aligned and the procedure for L3 and prisms alignment starts.
\begin{enumerate}
\item	The ADC -Corrector system will be dismounted the two prisms and the L3 lens will be place inside the ADC barrel and the system mounted back;
\item	The laser will be re aligned, but this time passing through the corrector lenses (not the ADC), when the laser is aligned the beam in principle is not deviated by the L1 and L2, therefore we can use the 2CCDs procedure described before and we will measure, as we planned for the L1 and L2, the position of the reflected beam and we will compute the shim size to be inserted;
\item	The ADC -Corrector system will be dismounted and the shims will be inserted on the side of the L3 and of the two prisms;
\item	The ADC-Corrector is mounted back;
\item	Laser aligned as described at the step 2;
\item	ADC barrel placed in position;
\item	Alignment will be verified for the ADC barrel lenses;
\item	OPTIONAL: Alignment will be verified for the Corrector barrel lenses again.
\end{enumerate}
After this operations the ADC and the corrector are aligned to the telescope optical axis.
\section{Conclusions}
In this paper we described the procedure we plan to follow at the end of this year (according to the last schedule, 2010) for the alignment of the optical and mechanical components of the VST. We designed several opto-mechanical tools to optimally accomplish this goal. The tools are ready and wait to be tested on the telescope. In these last lines we recall that the procedure described above is purely optical and relies only in the precision of the optical axis definition: in our case the instrument bearing of the telescope. However wobble and run-out of the bearing had been already measured with the bearing mount on the M1 cell before the shipment to Paranal Observatory and are fulfilling the (optical) specification we need.

\acknowledgments     
Since this paper offers an overview of the final step of the building of the VST telescope with the final positioning of the optical elements, the actual work was made by the INAF-Osservatorio di Capodimonte team, among them we remember M. Capaccioli, P. Schipani, S. d'Orsi, M. Brescia, L. Marty and D. Fierro.
We would like to say thanks to the ESO J. Spyromilio and R. Tamai for useful and meaningful discussions.


\bibliography{spie}   

\begin{thebibliography}{1}

\bibitem{2005SPIE.5962..608M}
{Marra}, G., {Mancini}, D., {Cortecchia}, F., {Sedmak}, G., and {Capaccioli},
  M., ``{VST optics design strategy and foreseen performance from U to I
  bands},'' {\em Proc. SPIE} {\bf 5962},  608--618 (2005).

\bibitem{2003SPIE.4837..379M}
{Mancini}, D., {Mancini}, G., {Perrotta}, F., {Ferragina}, L., {Fierro}, D.,
  {Fiume Garelli}, V., {Pellone}, L., {Caputi}, O., {Sciarretta}, G., and
  {Valentino}, M., ``{VST project: mechanical design optimization},'' {\em
  Proc. SPIE} {\bf 4837},  379--388 (2003).

\bibitem{2002Msngr.110...15K}
{Kuijken}, K., {Bender}, R., {Cappellaro}, E., {Muschielok}, B., {Baruffolo},
  A., {Cascone}, E., {Iwert}, O., {Mitsch}, W., {Nicklas}, H., {Valentijn},
  E.~A., {Baade}, D., {Begeman}, K.~G., {Bortolussi}, A., {Boxhoorn}, D.,
  {Christen}, F., {Deul}, E.~R., {Geimer}, C., {Greggio}, L., {Harke}, R.,
  {H{\"a}fner}, R., {Hess}, G., {Hess}, H., {Hopp}, U., {Ilijevski}, I.,
  {Klink}, G., {Kravcar}, H., {Lizon}, J.~L., {Magagna}, C.~E., {M{\"u}ller},
  P., {Niemeczek}, R., {de Pizzol}, L., {Poschmann}, H., {Reif}, K.,
  {Rengelink}, R., {Reyes}, J., {Silber}, A., and {Wellem}, W., ``{OmegaCAM:
  the 16k{$\times$}16k CCD camera for the VLT survey telescope},'' {\em The
  Messenger}~{\bf 110},  15--18 (2002).

\bibitem{2004SPIE.5492..513D}
{Diolaiti}, E., {Farinato}, J., {Ragazzoni}, R., {Vernet}, E., {Arcidiacono},
  C., and {Faccin}, F., ``{Optical alignment of the LBT prime focus camera},''
  {\em Proc. SPIE} {\bf 5492},  513--524 (2004).

\bibitem{2008SPIE.7014E.159G}
{Gentile}, G., {Ragazzoni}, R., {Diolaiti}, E., {Farinato}, J., {Hill}, J.,
  {Bertam}, R., and {Baruffolo}, A., ``{LBT report activities concerning the
  optomechanics alignment of the Large Binocular Camera's Red Channel},'' {\em
  Proc. SPIE} {\bf 7014} (2008).

\end{thebibliography}
\bibliographystyle{spiebib}   

\end{document}